\documentclass[fleqn,twoside]{article}

\usepackage{amsmath}
\usepackage{cite}
\input{xy}
\xyoption{all}

\makeatletter
 \def\@problemhead#1#2#3{%
  \par\kern-\parskip\kern#1
  \vbox\bgroup
  \hbox to\hsize{\hrulefill\raise1pt\hbox{\fbox{$\mathstrut$ #3\ }}\hrulefill}%
  \kern#2\par\kern-\parskip
 }
 \def\@problemtail#1#2{%
  \par\kern-\parskip\kern#1
  \hbox to\hsize{\hrulefill}
  \egroup
  \kern#2\par\kern-\parskip
 }

\makeatother

\usepackage{amsmath,amssymb}
\usepackage{espcrc2}

\usepackage{graphicx}

\title{
\vspace*{-5mm} \noindent
{\tiny \begin{flushleft} DESY 10--104 \hfill 
              \end{flushleft}}
Parton Distributions for LHC}

\author{S. Alekhin\address{DESY, Platanenalle 6, Zeuthen 15738, Germany}%
        \thanks{Permanent address: Institute for High Energy Physics, 
        Pobeda 1, Protvino 142281 Russia}
}
\begin{document}

\begin{abstract}
I review the current status of the nucleon PDFs determined from  
global fits with emphasis on the impact of recent experimental data
and the remaining theoretical challenges. 
\vspace{1pc}
\end{abstract}

\maketitle

\begin{table*}[htb]
\caption{Recently published nucleon PDFs sets with a brief description of the 
data used in the fit, the theoretical accuracy and the factorization 
scheme employed to model the heavy-quark DIS contribution.}
\label{table:1}
\newcommand{\m}{\hphantom{$-$}}
\newcommand{\cc}[1]{\multicolumn{1}{c}{#1}}
\renewcommand{\tabcolsep}{1.5pc} 
\renewcommand{\arraystretch}{1.2} 
\begin{tabular}{@{}lllll}
\hline
Name &  Data used & QCD approximation & Scheme & Reference \\  \hline

MSTW &  DIS+DY+jets & NLO/NNLO & GMVFN & \cite{MSTW} \\
CTEQ &  DIS+DY+jets & NLO & GMVFN & \cite{CTEQ} \\
NNPDF &  DIS+DY+jets & NLO & ZMVFN & \cite{NNPDF} \\
JR &  DIS+DY & NLO/NNLO & FFN & \cite{JR} \\
ABKM &  DIS+DY & NLO/NNLO & FFN & \cite{ABKM} \\
HERAPDF &  DIS & NLO & GMVFN & \cite{HERACOMB} \\

\hline
\end{tabular}\\[2pt]
\label{tab:pdfs}
\end{table*}

The detailed knowledge of 
the momentum distributions of partons in the nucleon is quite necessary  
for phenomenology of the hard scattering processes. Since the time, when  
the parton model was established
the parton distribution functions (PDFs) were permanently in the focus of 
the QCD studies. With a tremendous upgrade of the Tevatron 
collider luminosity and the start-up of the LHC at a record collision energy 
we need, however, an even better understanding of the PDF details in 
order to meet the improving accuracy of the experimental data. 
The PDFs are extracted from the hard-scattering 
data with the QCD evolution taken into account. At the present accuracy 
of the experimental data, moreover for the foreseen precision at the LHC, 
the QCD corrections at least up to the next-to-leading order (NLO) are
required. Furthermore, the next-to-next-to-leading order (NNLO) 
corrections are necessary for many important processes, as 
$W,Z$-bosons, Higgs boson and top-quark production. 

The choice of the data used to constrain the PDFs 
is often defined by the accuracy of the available theoretical calculations. 
For example, in order to include  the DIS data with low momentum 
transfers $Q^2$ into the PDF fit,
one has to take into account the higher-order QCD corrections and 
to take care about 
the higher-twist terms for the small-mass final state kinematics. Likewise, 
modifications of the nucleon wave function in 
nuclei have to be taken into account once nuclear data are used 
in the analysis.
\begin{figure}[t]
\includegraphics[width=17pc]{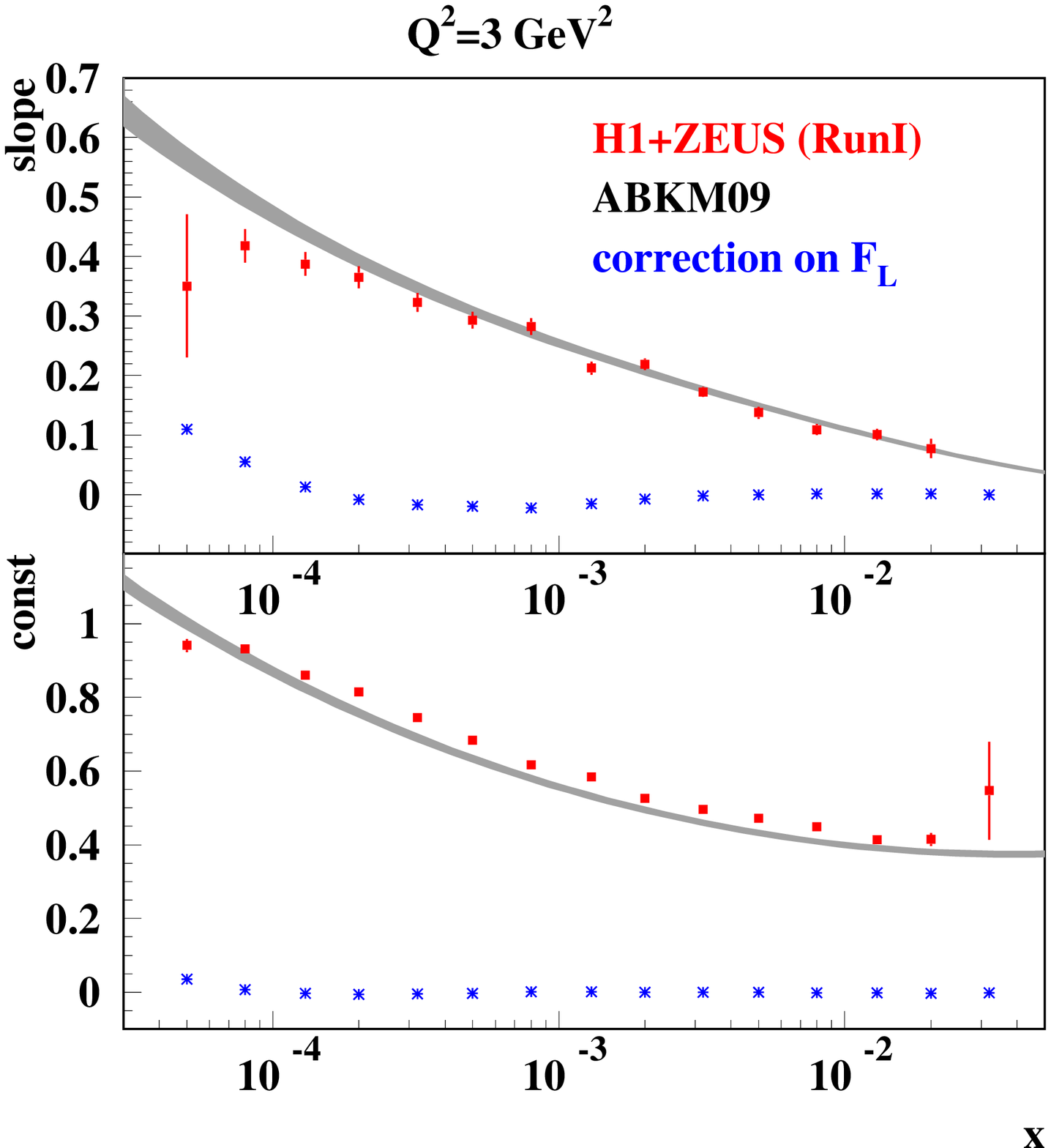}
\caption{The slope $d F_2/d \ln Q^2$ (upper panel) 
and the constant term (lower panel)
for the combined HERA Run~I data
on the inclusive SF $F_2$ in comparison to the predictions of the
ABKM09 fit~\cite{ABKM}. The impact of the correction on the
contribution of the longitudinal
SF $F_{\rm L}$, which were
 employed to extract the values of $F_2$ from the
cross sections data is given by stars. Figure taken from Ref.~\cite{ABKM-DIS}.
}
\label{fig:slope}
\end{figure}

\begin{figure}[t]
\includegraphics[width=17pc,height=15pc]{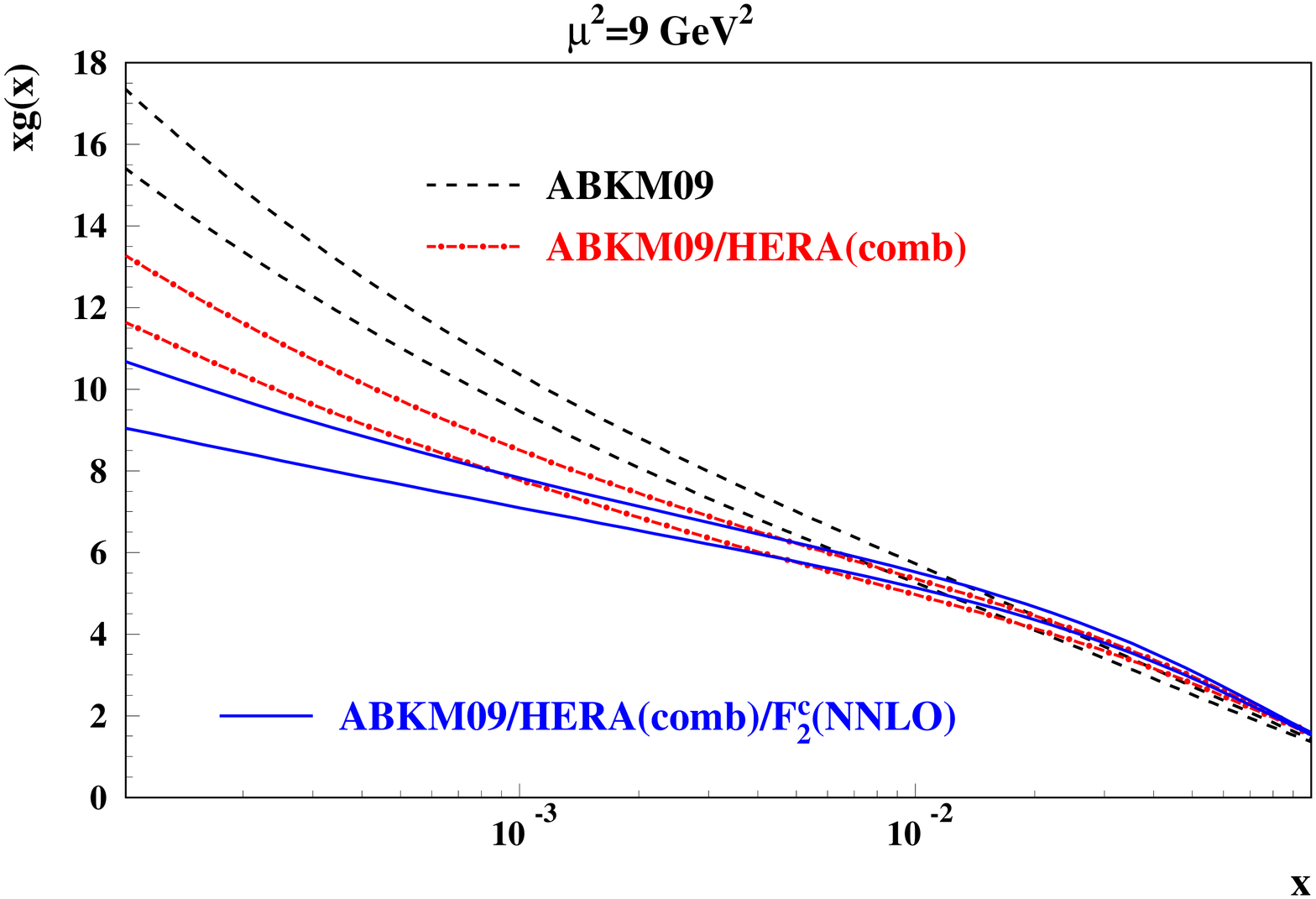}
\caption{The $1\sigma$ bands for the NNLO gluon distribution 
obtained in the updated version of our fit (blue) compared to 
the same for ABKM09 fit (black) and the version of the updated fit 
without partial NNLO corrections to the heavy-quark electro-production 
of Ref.~\cite{AM}
taken into account (red) at the factorization scale of $9~{\rm GeV}^2$.
Figure taken from Ref.~\cite{ABKM-DIS}.
}  
\label{fig:pdfs}
\end{figure}

There are several 
nucleon PDF sets, which were kept updated during the recent years, cf. 
Table~\ref{tab:pdfs}. All these PDF sets 
are based on the inclusive deep-inelastic-scattering (DIS) data collected 
in fixed-target experiments and the data from the HERA collider.
The HERA data mostly provide constraints on the gluon and sea quarks 
distributions at small values of the Bjorken variable $x$, 
which are particularly important
for the interpretation of the future data of the
first LHC run. Recently the 
inclusive data obtained by 
the H1 and ZEUS experiments of Run~I of the HERA collider were 
combined into the common data set~\cite{HERACOMB}. The advantage of such 
an approach is mutual cross-calibration of both experiments, which leads
to an essential improvement in the data accuracy. The slope 
of the updated HERA data on the inclusive structure 
function (SF) $F_2$ with respect to $\ln(Q^2)$, 
obtained from a model-independent fit, is given in Fig.~\ref{fig:slope}.
It is in good agreement with the NNLO predictions 
based on the ABKM09 PDFs~\cite{ABKM-DIS}. 
However, the constant term obtained in the 
same fit systematically overshoots the predictions. The 
combined HERA data in general lie above the separate data sets
of Ref.~\cite{H1ZEUS} used in the PDF fits before. This change in the data
requires a modification of the small-$x$ shape of the PDFs. In particular, 
for the version of the ABKM09 fit based on the combined HERA data, the 
gluon distribution at small $x$ is shifted
to lower values, cf. Fig.~\ref{fig:pdfs}. In the updated version of 
the NLO fit by the CTEQ collaboration a decrease of the 
gluon distribution at small $x$
is also observed~\cite{CTEQ}. However, the MSTW group found 
much smaller variations of the PDFs~\cite{MSTW-DIS}.

\begin{figure}[h]
\includegraphics[width=17pc,height=15pc]{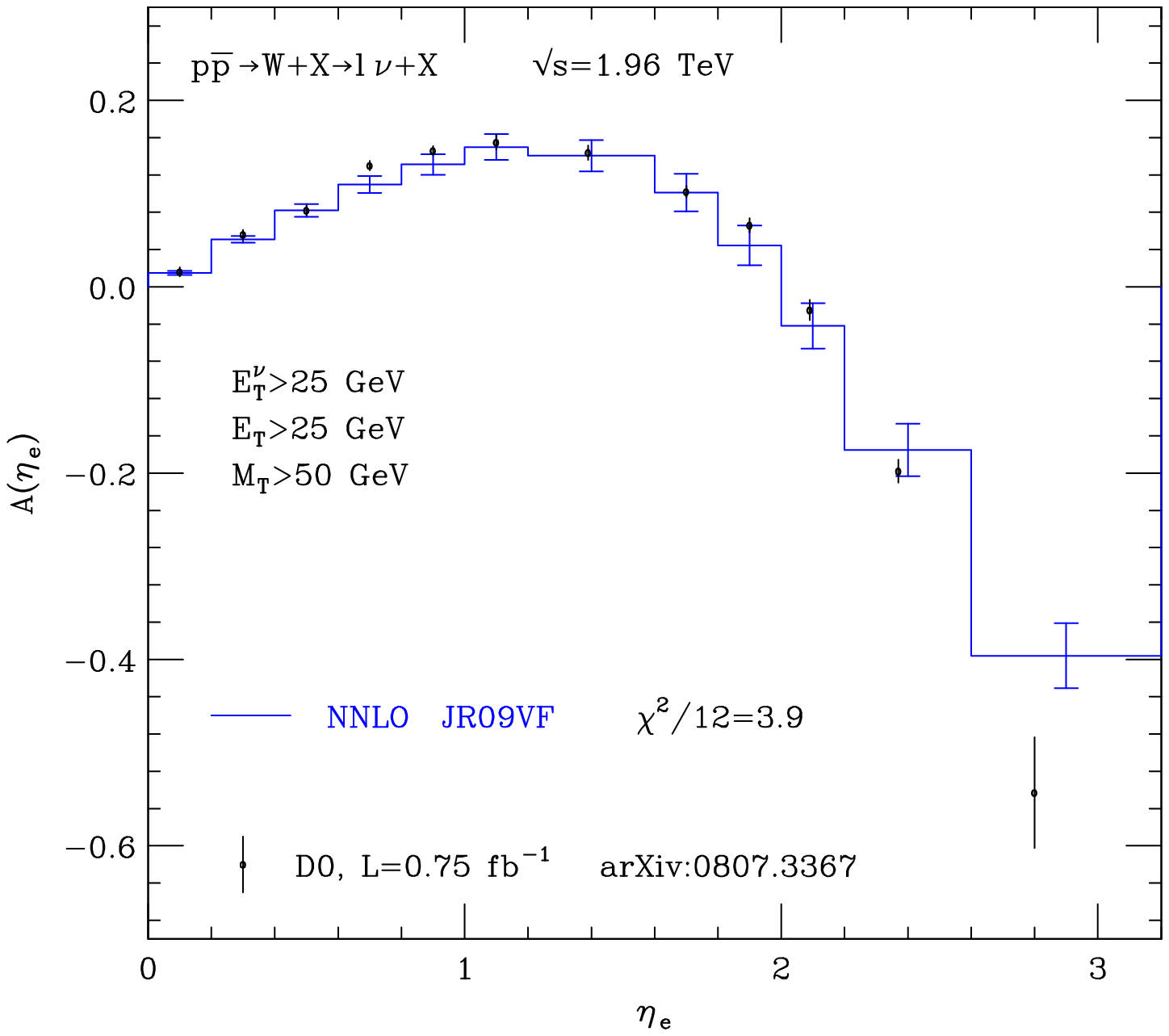}
\caption{The Run~II electron charge asymmetry obtained by the D0
collaboration in comparison with the NNLO predictions based on the JR 
PDFs~\cite{JR}. Figure taken from Ref.~\cite{Grazzini}. 
} 
\label{fig:lepasym}
\end{figure}

The PDF sets of Refs.~\cite{MSTW,CTEQ,NNPDF} were fitted also to the 
data on the $W^\pm$ and $Z$ production from hadron colliders. 
These collider data provide important constraints on the PDFs at 
large values of the factorization scale. The $W^\pm$, $Z$ rates
and distributions are calculated at NNLO~\cite{Melnikov,Grazzini} 
and are considered as one of the LHC luminosity monitors. 
However, the most recent and accurate Run~II Tevatron data~\cite{RUNII}
on the $W$ 
charge asymmetry and the related data on the lepton charge asymmetry are not 
in a good agreement with the NNLO calculations of Ref.~\cite{Grazzini} 
based on  
the different PDFs, particularly for the case of the electron charge asymmetry
data. The best agreement was observed for the case of 
the JR PDFs of Ref.~\cite{JR}, cf. Fig.~\ref{fig:lepasym}. 
The MSTW and ABKM09 predictions overshoot the data in general. The NNPDF
collaboration also obtained a poor description of the $W$-asymmetry data 
in their NLO fit~\cite{NNPDF}. Due to 
the QCD evolution of the PDFs the kinematics of the $W$-production 
at the Tevatron
is constrained by the fixed target DIS data at $x= 0.1\div 0.2$. Therefore 
the $W$-asymmetry predictions are mostly sensitive to the 
valence quark isospin asymmetry, $(u-d)$. This distribution 
can be constrained 
by the data on isospin-asymmetric combination of the proton and the neutron 
DIS SFs. However, the neutron 
target is available in a form of a bound state only. 
Therefore the neutron 
SFs unfolding requires to take into account nuclear effect. 
The DIS nuclear corrections
are non-negligible even for the lightest nuclei, as the deuteron.
Furthermore, they cannot be fully calculated using present
theoretical methods in nuclear physics and are often fitted to 
nuclear data instead, cf. e.g. Ref~\cite{KP}. The MSTW group included into 
their fit the deuteron correction  in a form of a 
model-independent function and fitted it to the global data 
set~\cite{MSTW-DIS}. In this way they found marginal improvement for 
the charged-lepton asymmetry. Moreover, a reasonable nuclear 
model cannot justify the form of the deuteron correction obtained.
Besides, the data on the lepton asymmetry
obtained by two Fermilab experiments, CDF and D0, are not in agreement.
In view of these problems the CTEQ group suggests two variants of their 
recent PDF release, with and without Run~II Tevatron data 
included~\cite{CTEQ}.

\begin{figure}[t]
\includegraphics[width=17pc,height=15pc]{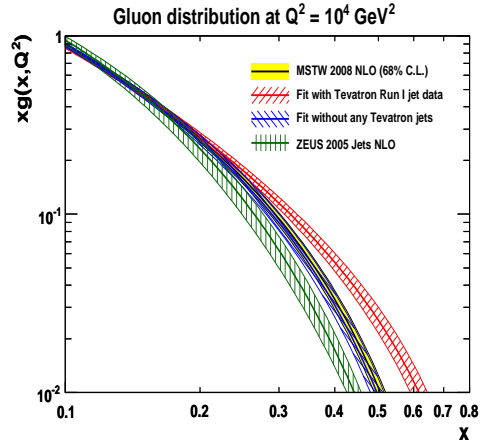}
\caption{The large-$x$ gluon distribution obtained in different variants 
of the NLO MSTW08 fit. Figure taken from Ref.~\cite{MSTW}.
}
\label{fig:gluon}
\end{figure}

The Tevatron jet inclusive production data were also 
used in Refs.~\cite{MSTW,CTEQ,NNPDF}. Historically, these data were always
considered as a unique source of the information about the gluon distribution 
at large $x$. Since the NNLO corrections  to the 
hadronic jet production are not yet known
these data can be used in a fully consistent way for extraction of 
the NLO PDFs only. With the Run~I data of Ref.~\cite{RUNIjet} 
included into the PDF
fit, the CTEQ and MSTW groups observed a substantial increase of the 
large-$x$ gluons~\cite{MRST,CTEQ4}. In fact, this effect was a manifestation 
of the tension between the Run~I jet data and the DIS data at large $x$. 
The accuracy of the data obtained in 
the Run~II experiments has been greatly improved, in particular due to the 
better jet energy calibration. Furthermore, these data lie lower than those  
from Run~I and are in much better agreement with the other data 
used in the PDF fits. 
As a result, the impact of the Run~II results on the PDFs is reduced, 
cf. Fig.~\ref{fig:gluon}.
Due to this change of trend, the value of the strong coupling 
constant $\alpha_{\rm s}$ extracted from the Tevatron data is shifted downward.
The shift in the value of $\alpha_{\rm s}$ is essential since 
in leading order QCD the 
jet production cross section is $O(\alpha_{\rm s}^2)$. 
The value of $\alpha_{\rm s}(M_Z)=0.1161^{+0.0011}_{-0.0018}$ 
was obtained in the recent analysis of the D0 data performed at the 
NLO with account of NNLO threshold corrections~\cite{Sonnenschein:2010tc}. 
Taking into account 
that the $\alpha_{\rm s}^{NNLO}$ is smaller than $\alpha_{\rm s}^{NLO}$,
this result is in a good agreement with the 
value of $\alpha^{NNLO}_{\rm s}(M_Z)=0.1147(12)$ 
obtained in the updated version of the ABKM fit~\cite{ABKM-DIS} and
with 
$\alpha^{N3LO}_{\rm s}(M_Z)=0.1141^{+0.0020}_{-0.0022}$ obtained in the 
analysis of the non-singlet DIS data~\cite{Blumlein:2006be}. 
The value of $\alpha^{NLO}_{\rm s}(M_Z)=0.1215$
obtained in the updated version of the MSTW fit~\cite{MSTW-DIS}   
is much larger than both these determinations.  
For comparison, the value of $\alpha_{\rm s}$ extracted from the $e^+e^-$
data is $\sim 0.119$~\cite{Bethke:2009jm}. However, it is worth to note 
that in the elaborated analysis of the global $e^+e^-$
data on the thrust distribution performed in the NNLO approximation 
with account of the power corrections to fragmentation 
the value of $\alpha_{\rm s}(M_Z)=0.1135\pm(0.0002)_{\rm exp.}
\pm(0.0005)_{\rm hadr.}\pm(0.0009)_{\rm pert.}$ was 
obtained~\cite{Abbate:2010xh}. This is much lower 
than the average of Ref.~\cite{Bethke:2009jm} and in good 
agreement with the determinations based on the DIS data. 

\begin{figure}[t]
\includegraphics[width=17pc,height=16pc]{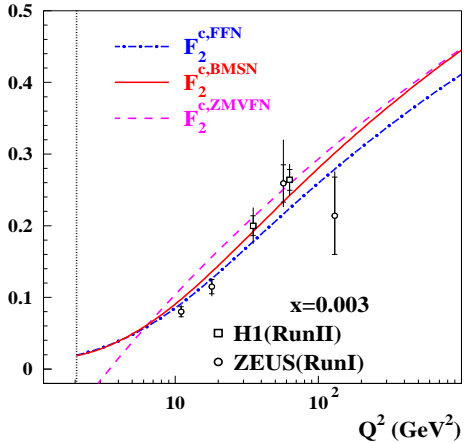}
\caption{ Comparison of $F_2^c$ in different 
    schemes to  H1- and ZEUS-data. 
    Solid line: GMVFN scheme in the 
    BMSN prescription, dash-dotted line: FFN scheme, dash line:
    ZMVFN scheme. Figure taken from Ref.~\cite{ABKM}.
}
\label{fig:f2c}
\end{figure}

The H1 and ZEUS data on the DIS charm production obtained in  
Run~I of the HERA collider have also been combined  into a common data  
set~\cite{HERACOMBCC} recently. 
Heavy-quark electro-production is particularly sensitive to the 
shape of the gluon distribution at small $x$. Therefore it gives a 
complimentary constraint for the global PDF fits. Moreover, the detailed 
understanding of the heavy-quark DIS electro-production mechanism is 
necessary for the small-$x$ inclusive SF description. 
In the global PDF fits the zero-mass
variable-flavor-number (ZMVFN) scheme is 
often used to model the heavy-quark contribution. 
In this scheme all quarks are assumed to be massless, which greatly 
simplifies the calculation. However, the low-$Q^2$ part of the DIS SFs 
cannot be calculated within this concept 
due to the power corrections and different constant terms occurring in the 
coefficient functions. The ZMVFN scheme commonly overestimates the 
heavy-quark contribution at small $Q^2$. Therefore it cannot be employed 
for the analysis of realistic DIS data. Instead of this the 
general-mass variable-flavor-number (GMVFN) scheme is used. 
The GMVFN schemes include a peculiar modeling of the low-$Q^2$ DIS region.
This allows to overcome problems of the ZMVFN scheme. However, since 
the modeling cannot be performed on solid theoretical grounds
the GMVFN scheme cannot be defined in a unique way. 
Various variants of 
the GMVFN scheme were considered, cf. Ref.~\cite{TUNG} and references 
therein. Further modifications of the ACOT scheme~\cite{Nadolsky:2009ge} 
and the updated version 
of Thorne's prescription~\cite{Thorne} were suggested.
The latter assumes a very flexible form of the semi-inclusive 
heavy-quark SF $F_2^h$ at low $Q^2$, 
which is provided by several 
additional parameters. An additional 
flexibility of model~\cite{Thorne} 
would evidently allow better description of the data 
within this approach. However, additional parameters introduced in the 
GMVFN schemes to model the DIS SFs at low $Q^2$ have to keep 
the factorization scheme consistency. Otherwise the PDFs obtained in the 
fit based on the GMVFN scheme might correspond to some peculiar factorization 
scheme, but not to the $\overline{\rm MS}$ scheme, commonly used in the QCD 
calculations, cf. Ref.~\cite{Chuvakin:1999nx}. In contrast, 
the BMSN prescription of the GMVFN scheme~\cite{BMSN} 
gives a consistent $\overline{\rm MS}$
description of the heavy-quark contribution in the transition region. 
At the same time it provides a smooth transition 
between the FFN scheme at low $Q^2$ and the ZMVFN scheme
at large $Q^2$, cf. Fig.~\ref{fig:f2c}.
These advantages of the BMSN ansatz were also 
employed in the FONLL approach to DIS considered in 
Ref.~\cite{Forte:2010ta}.

\begin{figure}[t]
\includegraphics[width=17pc,height=14.5pc]{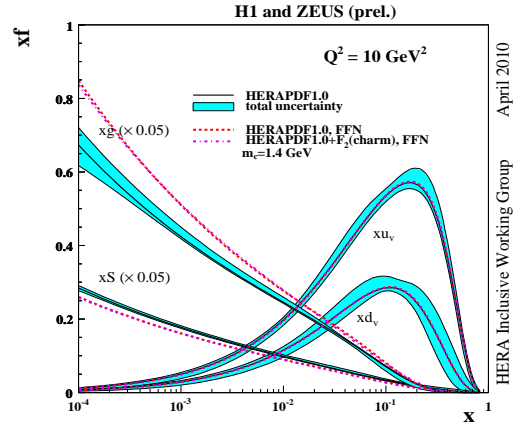}
\caption{Comparison of PDFs obtained in Ref.~\cite{HERACOMB}
for the GMVFN scheme of Ref.~\cite{Thorne} and in the variant of this fit 
for the FFN scheme. Figure taken from Ref.~\cite{HERACOMBCC}.
}
\label{fig:ffn}
\end{figure}
In the fixed-flavor-number (FFN) scheme
the heavy-quark mass dependence follows from a complete fixed-order 
computation without any modeling even at low $Q^2$.
The coefficient functions of the heavy-quark semi-inclusive SFs are calculated 
in complete form up to the $O(\alpha_{\rm s}^2)$ (NLO)~\cite{Laenen:1991af}
semianalytically. 
At the values of $Q^2\gg m_h^2$ the fixed-order FFN expressions are  
insufficient since they do not include the big-log terms of 
$O(\alpha_{\rm s}^n\ln ^m(Q^2))$, which might be important even 
for large $n$. In the ZMVFN scheme these terms are resummed in a natural way 
through the QCD evolution of the heavy-quark PDFs. This is often considered 
as an advantage of the ZMVFN schemes as compared to the FFN one. Nonetheless, 
these terms are partially taken into account in the NLO expressions of   
Ref.~\cite{Laenen:1991af} and the numerical effect of the remaining 
big-log terms is marginal for the realistic data kinematics.
This observation was first made in Ref.~\cite{Gluck:1993dpa} and later 
confirmed in Ref.~\cite{ABKM} for the case of recent data on $F_2^c$ and 
$F_2^b$ accumulated by the HERA experiments.
In particular, the difference 
between the PDFs obtained in the FFN fit and in the GMVFN fit based on the BMSN 
prescription was found to be marginal~\cite{ABKM}.
 For the PDF fit, which includes the combined HERA data on 
$F_2^c$, the FFN scheme and the different variants of the 
GMVFN schemes provide equally good description of the data, 
cf. Ref.~\cite{HERACOMBCC}. 
At the same time the gluon distributions obtained for the 
GMVFN scheme of Ref.~\cite{Thorne} and the FFN PDFs are quite different, 
particularly 
for the small-$x$ gluons, cf. Fig.~\ref{fig:ffn}. The ZMVFN gluons at small 
$x$ must take lower values 
than for the FFN scheme by the scheme definition. However, this 
does not fully explain the effect given in Fig.~\ref{fig:ffn}.
Meanwhile this effect might be related to
the discrepancy in the gluon distributions 
obtained in the fits of Ref.~\cite{MSTW} and Ref.~\cite{ABKM}.

\begin{figure}[t]
\includegraphics[width=17pc,height=18pc]{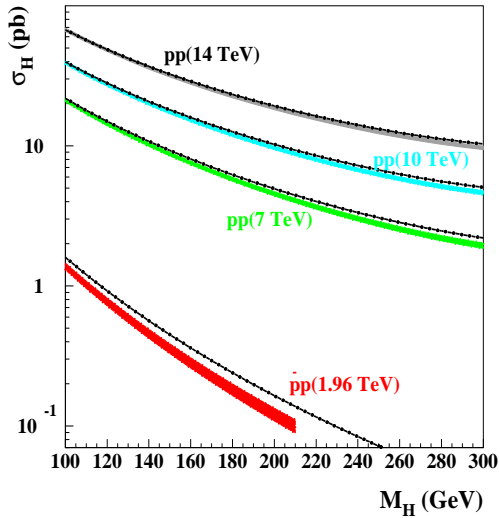}
\caption{The $1\sigma$ error band for 
the Higgs-boson production cross sections 
at Tevatron and the LHC employing the ABKM PDFs~\cite{ABKM}
in comparison with the central values for the case 
of MSTW PDFs~\cite{MSTW} (dash-dotted lines).
 Figure taken from Ref.~\cite{ABKM}.
}
\label{fig:higgs}
\end{figure}
The gluon-gluon collision is the dominant mechanism for the production of  
Higgs bosons with masses $100\lesssim M_{\rm H}\lesssim 300~{\rm GeV}$  
in hadron-hadron collisions. Furthermore, the leading order 
gluon-gluon term is $O(\alpha^2_{\rm s})$. This makes the 
Higgs production rate very sensitive to the variation of the gluon shape and 
the value of $\alpha_{\rm s}$. The Higgs production rates at the   
Tevatron collider and at LHC energies calculated with account 
of the NNLO corrections of Ref.~\cite{Harlander:2002wh} are given 
in Fig.~\ref{fig:higgs}. The estimates based on the NNLO ABKM09 PDFs take
somewhat lower values than those  based on the MSTW08 PDFs, particularly 
for the case of Tevatron. This is explained by a cumulative effect of the 
difference in the gluon distribution and $\alpha_{\rm s}$ for these 
two sets. This discrepancy reduces the statistical significance 
of the constraint on the Higgs mass obtained at the Tevatron collider,
cf. Refs.~\cite{Aaltonen:2010yv,Baglio:2010um}. For the 
conclusive interpretation of the upcoming LHC results
it would be useful to reduce the spread of the predictions. 

In summary, the ensemble of nucleon PDF sets is maintained at the moment and 
is ready for phenomenological studies at the LHC. The PDFs are tuned to 
a variety of the hard-scattering data, including the 
most recent ones,  and provide an accuracy 
of the LHC standard candle processes at the level of several 
percent~\cite{LHC}. Nonetheless the predictions based on the different PDFs 
are not in perfect agreement, presumably due to the distinct details 
in the theoretical treatment of the data employed by different groups. 
Clarification of these aspects and consolidation of the predictions   
will essentially improve the precision of the future LHC findings.

{\bf Acknoledgments.}{ I would like to thank J.Bl\"umlein for careful 
reading of the manuscript and valuable comments. 
This work was supported in part by Helmholtz Alliance
``Physics at the Terascale''.}

\end{document}